\UseRawInputEncoding    
\documentclass{ws-mpla}
\usepackage[super]{cite}
\usepackage{graphicx}

\newcommand{\xv}{\vec{\rm x}}

\newcommand{\sv}{\vec{\rm s}}

\newcommand{\nv}{\vec{\rm n}}
\newcommand{\Av}{\vec{\rm A}}
\newcommand{\ev}{\vec{\rm e}}

\newcommand%
{\MVV}%
[1]%
{{\langle\!\langle #1\rangle\!\rangle}}%
\newcommand%
{\MMV}%
[1]%
{{\left\langle\!\!\!\left\langle #1\right\rangle\!\!\!\right\rangle}}%
\def\bldmth#1{%
\mathchoice
{{\hbox{\boldmath$\displaystyle#1$\unboldmath}}}%
{{\hbox{\boldmath$\textstyle#1$\unboldmath}}}%
{{\hbox{\boldmath$\scriptstyle#1$\unboldmath}}}%
{{\hbox{\boldmath$\scriptscriptstyle#1$\unboldmath}}}%
}
\def\vec#1{\bldmth{#1}}
\begin{document}

\markboth{S.~E.~Korenblit and E.~A.~Voronova}
{Non-abelian monopole solutions in arbitrary gauge}

\catchline{}{}{}{}{}

\title{Non-abelian monopole solutions in arbitrary gauge}

\author{\footnotesize S.~E.~Korenblit${}^{\dagger,*}$ and E.~A.~Voronova${}^\dagger$}

\address{
${}^\dagger$ Department of Physics, Irkutsk State University, \\
20 Gagarin blvd, Irkutsk 664003, Russia  \\
${}^*$ Dzhelepov Laboratory of Nuclear Problems, Joint Institute for Nuclear Research, \\ 
RU-141980, Dubna, Russia.  \\
${}^*$korenb@ic.isu.ru  $ \quad {}^\dagger$vreaj@mail.ru 
}

\maketitle

\begin{abstract}
The Georgi-Glashow model equations of motion are examined by general static spherically 
symmetric real and complex parametrizations of gauge fields in arbitrary gauge.
Their connection with the known `t Hooft-Polyakov and Julia-Zee equations is shown without 
any gauge-fixing, elucidating so the gauge invariant meaning of corresponding functions. 
The obtained systems of equations have a new exact analytical solutions for complex 
non-abelian monopole and dyon fields with real finite or zero energy densities.  

\keywords{monopoles, dyons, gauge conditions, self-dual fields, dark energy}
\end{abstract}

\ccode{PACS Nos.: 11.15.Ex; 14.80.Hv;}

\section{Introduction}
The seminal conception of abelian Dirac monopole today is 90 years old 
\cite{Dr,Schw,Schw_m,Milt,Kr_Li}. It plays an important role in a number of field 
theoretic considerations, where it can account for the quantization of the electric charge.  
It has got its second life in non-abelian gauge theories with (or without) spontaneous 
symmetry breaking according to Higgs mechanism \cite{G-G}. It was shown that these theories 
naturally incorporate the Dirac monopole as abelian projection of their non-abelian monopole 
solutions \cite{Pol,tH,Ar,Blw,JZ,Grac,Rbk,Gd_Ol,Wpf,YS,Svrz,MPr,Bgm,Grs,Zng,Corrig}, which,  
in turn, can catalyze the decay of proton \cite{Rubak} or provide quarks confinement 
\cite{Cho_13}. 

For the stationary case with $\partial^0\mapsto 0$, where $x^\mu=(x^0,\xv)$, $\xv=r\nv$ for  
$r=|\xv|$, the time independence presupposes a choice of Lorentz frame in which the fields are 
at rest \cite{Gd_Ol}. In the rest frame of this classical field the absence of 
``chromo-electric'' field strength $E^j_a(x)=0$ implies the temporal gauge condition to 
non-abelian gauge potentials $A^\mu_a(x)=(A^0_a(x),\Av_a(x))$ as $A^0_a(x)=0$. 
The well known classical static `t Hooft-Polyakov solution 
\cite{Pol,tH,Ar,Blw,JZ,Grac,Rbk,Gd_Ol,Wpf,YS} in fact satisfies an additional transverse  
gauge condition $\left(\nv\cdot\Av_a(x)\right)=0$ and Coulomb gauge condition \cite{JZ} 
$\left(\vec{\nabla}\cdot\Av_a(x)\right)=0$. The last condition excludes from the potential 
$A^j_a(x)$ not only the longitudinal structure $n^jn^a$ but also the transverse one 
$(\delta^{ja}-n^jn^a)$, inevitably accompanied it (clf. (\ref{2_13}), (\ref{2_14}) below).

In the following sections it is shown how the consistent removing of those gauge constraints 
extends and generalizes the interpretation of the functions of `t Hooft-Polyakov 
and Julia-Zee solutions and leads to new exact classical solutions for complex 
non-abelian monopole and dyon fields with real finite or zero energy densities.


\section{Georgi-Glashow model and `t Hooft-Polyakov ansatz} 
We shall start with reminding of group representation, covariant derivative $\widehat{D}^\mu$, 
field strength $G^{\mu\nu}_a$, Lagrangian, equations of motion, energy density and vacuum 
expectation for $SO(3)$ Georgi-Glashow model \cite{G-G}. This model contains the 
gauge potentials $A^\mu_a(x)$ and scalar Higgs fields $h_a(x)$ in adjoint representation for  
$\widehat{A}^\mu=A^\mu_a T^a$ and $\widehat{h}=h_a T^a$, that for 
$\partial^\mu=(\partial^0,-\vec{\nabla})$ and generators $T^a$ obey the relations 
\cite{Blw,JZ,Grac,Rbk,Gd_Ol,Wpf,YS}:  
\begin{eqnarray}
&&\!\!\!\!\!\!\!\!\!\!\!\!\!\!\!\!\!\!\!\!\!\!\!\! 
\left[T^a,T^b\right]=i\varepsilon^{abc}T^c,\quad 
\left(T^a\right)_{bc}=- i\varepsilon^{abc},\; \mbox{ -- in adjoint representation},  
\label{2_1} \\
&&\!\!\!\!\!\!\!\!\!\!\!\!\!\!\!\!\!\!\!\!\!\!\!\! 
\widehat{D}^\mu=\widehat{I}\partial^\mu -ie\widehat{A}^\mu, \quad 
\left(\widehat{D}^\mu\widehat{h}\right)_a\equiv \left[\widehat{D}^\mu,\widehat{h}\right]_a=
\partial^\mu h_a+e\varepsilon^{abc}A^\mu_b h_c,  
\label{2_2} \\
&&\!\!\!\!\!\!\!\!\!\!\!\!\!\!\!\!\!\!\!\!\!\!\!\! 
e \widehat{G}^{\mu\nu}=eG^{\mu\nu}_a T^a=i\left[\widehat{D}^\mu,\widehat{D}^\nu\right],
\quad 
G^{\mu\nu}_a=\partial^\mu A^\nu_a-\partial^\nu A^\mu_a+e\varepsilon^{abc}A^\mu_b A^\nu_c.  
\label{2_3}
\end{eqnarray}
Here $\widehat{I}$ is unit matrix in corresponding representation and $\hbar=c=1$.
The Lagrangian density, Euler equations, Bianchi identity, symmetric energy-momentum tensor 
$\Theta^{\mu\nu}(x)$, energy density and vacuum expectation read correspondingly as 
\cite{Blw,JZ,Grac,Rbk,Gd_Ol,Wpf,YS} 
\begin{eqnarray}
&&\!\!\!\!\!\!\!\!\!\!\!\!
{\cal L}(x)=-\frac{1}{4}G^{\mu\nu}_aG_{\mu\nu}^a+\frac{1}{2}
\left(\widehat{D}^\mu\widehat{h}\right)_a \left(\widehat{D}_\mu\widehat{h}\right)_a-
\frac{\lambda}{4}\left(h_ah_a-{\rm f}^2\right)^2,
\label{2_4} \\
&&\!\!\!\!\!\!\!\!\!\!\!\! 
\left(\widehat{D}_\nu  \widehat{G}^{\mu\nu}\right)_a\!=
e\varepsilon^{abc}h_b\left(\widehat{D}^\mu\widehat{h}\right)_c, \quad 
\left(\widehat{D}_\mu\widehat{D}^\mu\widehat{h}\right)_a\!=
-\lambda h_a\left(h_b h_b-{\rm f}^2\right),  
\label{2_5}  \\
&&\!\!\!\!\!\!\!\!\!\!\!\!
\varepsilon_{\mu\nu\sigma\varkappa}
\left(\widehat{D}^\nu\widehat{G}^{\sigma\varkappa}\right)_a\!\equiv 0,\quad 
\Theta^{\mu\nu}(x)=-\,G^{\mu\varkappa}_a G^{\nu}_{a\,\varkappa}+
\left(\widehat{D}^\mu\widehat{h}\right)_a\!\left(\widehat{D}^\nu\widehat{h}\right)_a\!-
\eta^{\mu\nu}{\cal L}(x), 
\label{2_6_0}  \\
&&\!\!\!\!\!\!\!\!\!\!\!\!
\Theta^{00}(x)=\frac 12\left\{\left(E^j_a\right)^2\!+\left(B^j_a\right)^2\!+
\left[\left(\widehat{D}^0\widehat{h}\right)_a\right]^2\!+
\left[\left(\widehat{D}^j\widehat{h}\right)_a\right]^2 \right\}+
\frac{\lambda}{4}\left(\left(h_b\right)^2-{\rm f}^2\right)^2\!, 
\label{2_6}  \\
&&\!\!\!\!\!\!\!\!\!\!\!\! 
\mbox{with }\; 
E^j_a=-G^{0j}_a,\quad B^l_a=-\,\frac 12 \varepsilon^{jkl}G^{jk}_a, \quad 
\langle 0|h_a(x)|0\rangle={\rm f}\delta^{a3}, \quad e^2{\rm f}^2=M^2. 
\label{2_7} 
\end{eqnarray}
Here $M$ is the mass acquired by the gauge fields $A^\mu_1,A^\mu_2$ after absorption of 
Higgs fields $h_1,h_2$ \cite{YS}. The remaining massless component $A^\mu_3$ corresponds to 
residual $U(1)$ gauge symmetry responding for existence of abelian monopole in this theory 
and is connected with residual freedom of rotation around third axis in isotopic (``color'') 
space, defined \cite{Grac,Rbk,Gd_Ol,Wpf,YS,Svrz,MPr} by vacuum expectation ${\rm f}$ of 
$h_3(x)$ in the unitary gauge (\ref{2_7}). 

The static `t Hooft-Polyakov solution for $A^0_a(\xv)=0$ with 
$\left(\nv\cdot\Av_a(\xv)\right)=0$, $\left(\vec{\nabla}\cdot\Av_a(x)\right)=0$ is defined
\cite{Pol,tH,Ar,Blw,JZ,Grac,Rbk,Gd_Ol,Wpf,YS} by two real functions $K(r)$ and $H(r)$ 
as\footnote{Here and below the mixed latin indexes of 3D- configuration space and isotopic 
(color group) space $j,k,l,s=1,2,3$ and $a,b,c,d=1,2,3$, are sum up by Euclidean metric 
everywhere also as Lorentz upper ones, whereas the Minkovski metric is usual 
$\eta^{\mu\nu}=diag(1,-1,-1,-1)$. Bold liters are used only for 3D vectors in configuration 
space: 
$\left(B^j_a\right)^2=\left(\vec{B}_a\right)^2=\left(\vec{B}_a\!\cdot\vec{B}_a\right)$, 
e.t.c.}: 
\begin{eqnarray}
&&\!\!\!\!\!\!\!\!\!\!\!\!\!\!\!\!\!\!
A^j_a(\xv)=-\varepsilon^{jka}\frac{n^k}{er}\left[1-K(r)\right], \qquad 
h_a(\xv)=n^a\frac{H(r)}{er} \equiv \pm  n^a {\rm f}\left[1-S(r)\right], 
\label{2_8} \\
&&\!\!\!\!\!\!\!\!\!\!\!\!\!\!\!\!\!\! 
G^{jk}_a(\xv)\equiv -\varepsilon^{jkl}B^l_a(\xv)=\frac{\varepsilon^{jkl}}{er^2}\left\{
n^l n^a \left[K^2(r)-1\right]+\left(\delta^{la}-n^l n^a\right)r\frac{dK(r)}{dr}\right\}. 
\label{2_9} 
\end{eqnarray}

\section{Spherically symmetric gauge fields} 
We consider the most general static spherically symmetric ({\bf ss}) ansatz for the gauge 
fields \cite{Rbk,Gd_Ol,Wpf,YS,Svrz} as a superposition of three indpendent {\bf ss}- 
structures, mutualy orthogonal in configuration and isotopic spaces simultaneously, 
for $\xv=r\nv$, with $\vec{\nabla}r=\nv$: 
\begin{eqnarray}
&&\!\!\!\!\!\!\!\!\!\!\!\!\!\!\!\!\!\!
A^j_a(\xv)=\varepsilon^{jka}n^k \frac{\gamma(r)}{er}-n^jn^a\frac{\alpha(r)}{er}-
\left(\delta^{ja}-n^j n^a\right)\frac{\beta(r)}{er},\;\,\mbox{ so that now:}  
\label{2_13} \\
&&\!\!\!\!\!\!\!\!\!\!\!\!\!\!\!\!\!\!
\left(\nv\cdot\Av_a(\xv)\right)=-n^a\frac{\alpha(r)}{er}, 
\qquad \left(\vec{\nabla}\cdot\Av_a(\xv)\right)=
\frac{n^a}{e r^2}\left[2\beta(r)-\partial_r\left(r\alpha(r)\right)\right], 
\label{2_14} 
\end{eqnarray} 
where the second longitudinal structure $n^jn^a$ also is orthogonal to the both transverse 
ones $\varepsilon^{jka}n^k$ and $\left(\delta^{ja}-n^j n^a\right)$ separately in configuration 
and isotopic spaces. Without the loss of generality, the gauge invariance of the theory allows 
to impose the conditions onto the function $\alpha(r)$ to be real and regular everywhere 
including infinity. The conditions for the functions $\beta(r),\gamma(r)$ will be given below.  

The conditions of spherical symmetry for {\bf ss}-- vector field (\ref{2_13}) as well as for  
scalar field (\ref{2_8}) mean \cite{Rbk,Gd_Ol,Wpf,YS,Svrz,MPr} their invariance under 
rotation ${\cal R}( {\rm g})$ in configuration space and global gauge transformation 
${\rm V}( {\rm g})$ simultaneously, with the same arbitrary rotation parameters from diagonal 
subgroup ${\rm g}\in SO(3)_{\rm d}$ of such simultaneous rotations in both configuration 
and isotopic spaces: 
\begin{eqnarray}
&&\!\!\!\!\!\!\!\!\!\!\!\!\!\!
{\rm V}( {\rm g})\!\left[\!{\cal R}( {\rm g})\widehat{\Av}\!
\left({\cal R}^{-1}( {\rm g})\xv\right)\!\right]\!{\rm V}^{-1}( {\rm g})=
\widehat{\Av}(\xv),\quad 
{\rm V}( {\rm g})\widehat{h}\!\left({\cal R}^{-1}( {\rm g})\xv\right)\!
{\rm V}^{-1}( {\rm g})=\widehat{h}(\xv), 
\label{2_15} 
\end{eqnarray} 
which holds for each of these three {\bf ss}- structures separately. These {\bf ss}-- 
structures define also the ``chromo-magnetic'' field (\ref{2_3}), (\ref{2_7}), 
where, unlike the potentials (\ref{2_13}), the longitudinal structure $n^jn^a$ appears as 
gauge (i.e. $\alpha$)- independent and the prime means the derivative with respect to $r$ also 
for the function $Y(r)=1+\gamma(r)$:  
\begin{eqnarray}
&&\!\!\!\!\!\!\!\!\!\!\!\!\!\!
B^j_a(\xv)=\frac{n^j n^a}{er^2}\!\left[1-Y^2-\beta^2\right]\!-
\frac{\left(\delta^{ja}-n^j n^a\right)}{er^2}\left(rY^\prime+\alpha\beta\right)-
\frac{\varepsilon^{jka}n^k}{er^2}\left(r\beta^\prime-\alpha Y\right)\!.
\label{2_16} 
\end{eqnarray} 
With the same ansatz (\ref{2_8}) for the Higgs field, by means of independence and 
orthogonality of the above three {\bf ss}- structures, the equations of motion (\ref{2_5}) 
lead to the following system of four equations, it would seem, for four unknown functions: 
\begin{eqnarray}
&&\!\!\!\!\!\!\!\!\!\!\!\!\!\!\!\!\!\! 
r^2\frac{d}{dr}\left(\frac{rY^\prime+\alpha\beta}{r}\right)=
Y\left(Y^2+\beta^2+H^2-1\right)-\alpha\left(r\beta^\prime-\alpha Y\right),
\label{2_17_1} \\
&&\!\!\!\!\!\!\!\!\!\!\!\!\!\!\!\!\!\! 
r^2\frac{d}{dr}\left(\frac{r\beta^\prime-\alpha Y}{r}\right)=
\beta\left(Y^2+\beta^2+H^2-1\right)+\alpha \left(rY^\prime+\alpha\beta\right),
\label{2_17_2} \\
&&\!\!\!\!\!\!\!\!\!\!\!\!\!\!\!\!\!\! 
\beta\left(rY^\prime+\alpha\beta\right)=Y\left(r\beta^\prime-\alpha Y\right), 
\label{2_17_3} \\
&&\!\!\!\!\!\!\!\!\!\!\!\!\!\!\!\!\!\! 
r^2\frac{d^2H}{dr^2}=H\left\{\frac{\lambda}{e^2}\left[H^2-(Mr)^2\right]+
2\left(Y^2+\beta^2\right)\right\}. 
\label{2_17_4}
\end{eqnarray}
It may be obtained also by variation with respect to the corresponding functions 
$Y,\beta,\alpha,H$ of the energy functional \cite{Svrz} for the energy density (\ref{2_6}) 
with the gauge field (\ref{2_13}) and strength (\ref{2_16}) (for corresponding stress tensor 
$\Theta^{jk}_m(\xv)$ see the Appendix, Eqs. (\ref{A_1}), (\ref{A_2}) with $\vec{E}_a=0$ and 
$J=0$.): 
\begin{eqnarray}
&&\!\!\!\!\!\!\!\!\!\!\!\!\!\!\!\!\!\! 
{\rm E}_m[\Theta]=\int d^3{\rm x}\,\Theta^{00}_m(x)\longmapsto 
4\pi \int\limits^\infty_0 r^2 dr\,\Theta^{00}_m(r)=
\label{2_18_1} \\
&&\!\!\!\!\!\!\!\!\!\!\!\!\!\!\!\!\!\! 
=\frac{4\pi}{e^2}\int\limits^\infty_0\frac{dr}{r^2}\left\{
\frac 12\left[1-Y^2-\beta^2\right]^2+\left(rY^\prime+\alpha\beta\right)^2+
\left(r\beta^\prime-\alpha Y\right)^2+\right.
\nonumber \\ 
&&\!\!\!\!\!\!\!\!\!\!\!\!\!\!\!\!\!\! 
\left. +\frac 12\left(rH^\prime-H\right)^2+H^2\left(Y^2+\beta^2\right)+
\frac{\lambda}{4e^2}\left[H^2-(Mr)^2\right]^2 \right\}. 
\label{2_18_3} 
\end{eqnarray}
Note that this quantity and its density $\Theta^{00}_m(r)$ being observables should be real 
and gauge (i.e. $\alpha$)- independent also for the complex unobservable gauge-dependent 
non-abelian gauge potentials (\ref{2_13}) and field strengths (\ref{2_16}). 

For $Y\neq 0$ and $Y^2+\beta^2\neq 0$ the third Eq. (\ref{2_17_3}) is immediately integrated 
for  
\begin{equation}
\frac{\beta(r)}{Y(r)}=\tan \omega(r),\;\, \mbox{ as }\;\,
r\omega^\prime(r)=\alpha(r),
\;\,\mbox{ and }\;\,\omega(r)=
\int\limits^r_{}\! d\xi\,\frac{\alpha(\xi)}{\xi} 
\quad ({\rm mod}\,\pi).  
\label{2_20} 
\end{equation}
Then the symmetry of the system (\ref{2_17_1})--(\ref{2_17_4}) under the simultaneous  
changing of $Y\mapsto\pm\widetilde{\beta}$ and $\beta\mapsto\mp\widetilde{Y}$ can be realised 
in the form: 
\begin{equation}
Y(r)=K(r)\cos\omega(r), \quad \beta(r)=K(r)\sin\omega(r),\;\mbox{ where }\; 
Y^2+\beta^2=K^2>0,  
\label{2_19} 
\end{equation}
with the real functions $K(r)$ and  $\omega(r)$, providing this symmetry for real $Y,\beta$ by 
the simple shifting $\omega\mapsto\widetilde{\omega}\mp\pi/2$. 
Then each of the first two Eqs. (\ref{2_17_1}), (\ref{2_17_2}) and the fourth 
Eq. (\ref{2_17_4}) are reduced exactly to the first and second `t Hooft-Polyakov equations 
for the functions $K(r)$ and $H(r)$ respectively, being the same as those for the 
`t Hooft-Polyakov ansatz (\ref{2_8}): 
\begin{equation}
r^2\frac{d^2K}{dr^2}=K\left[K^2+H^2-1\right], \quad 
r^2\frac{d^2H}{dr^2}=H\left\{\frac{\lambda}{e^2}\left[H^2-(Mr)^2\right]+2K^2\right\}. 
\label{2_10}
\end{equation} 
They are of course reproduced by the system of Eqs. (\ref{2_17_1})--(\ref{2_17_4}) for 
$\alpha=\beta=0$ and $Y\mapsto K$. The functions $\alpha,\,\omega$ describing the gauge 
arbitrariness (\ref{2_14}) could not be defined by any dynamical equations. 
Nevertheless, their inclusion demonstrates the gauge invariant meaning of `t Hooft-Polyakov 
functions $K(r)$ and $H(r)$. The gauge-dependent non-abelian ``chromo-magnetic'' field 
strength (\ref{2_16}) now reads as 
\begin{eqnarray}
&&\!\!\!\!\!\!\!\!\!\!\!\!\!\!
B^j_a(\xv)=\frac{n^j n^a}{er^2}\!\left[1-K^2\right]\!-
\Xi^{ja}(\nv,\omega) \frac{rK^\prime}{er^2}, 
\label{2_21} \\
&&\!\!\!\!\!\!\!\!\!\!\!\!\!\!
\mbox{with  }\;\;\,
\Xi^{ja}(\nv,\omega)=\left(\delta^{ja}-n^j n^a\right) \cos\omega+
\varepsilon^{jka}n^k \sin\omega, 
\label{2_21_1} 
\end{eqnarray} 
as transverse gauge-dependent tensor. For $\omega(r)=0$ this strength obviously returns to 
the `t Hooft-Polyakov strength (\ref{2_9}) as its natural gauge-dependent generalization. 

The full energy (\ref{2_18_3}) for the solution (\ref{2_19}), (\ref{2_21}) also takes its 
known form \cite{Rbk,Gd_Ol,Wpf,YS} independent of $\omega$, $\alpha$ 
(stress tensor $\Theta^{jk}_m(\xv)$ see in (\ref{A_4}), (\ref{A_6})--(\ref{A_9}) with $J=0$): 
\begin{eqnarray}
&&\!\!\!\!\!\!\!\!\!\!\!\!\!\!\!\!\!\! 
{\rm E}_m[\Theta]=\frac{4\pi}{e^2}\int\limits^\infty_0\frac{dr}{r^2}\left\{
\frac 12\left[1-K^2\right]^2+\left(rK^\prime\right)^2+ \right.
\nonumber \\ 
&&\!\!\!\!\!\!\!\!\!\!\!\!\!\!\!\!\!\! 
\left. +\frac 12\left(rH^\prime-H\right)^2+H^2K^2+
\frac{\lambda}{4e^2}\left[H^2-(Mr)^2\right]^2 \right\}. 
\label{2_18_4} 
\end{eqnarray}
The boundary conditions to nonlinear Eqs. (\ref{2_10}) is more subtle question 
\cite{Blw,JZ,Grac,Rbk,Gd_Ol,Wpf}. The asymptotics at $r\to\infty$ is strictly determined by 
topological features of monopole solutions 
\cite{Pol,tH,Ar,Blw,JZ,Grac,Rbk,Gd_Ol,Wpf,YS,Svrz,MPr} as vanishing of $K(r)$ and $S(r)$. 
With the help of this conditions one can linearize the Eqs. (\ref{2_10}) 
at $r\to\infty$ to obtain \cite{Gd_Ol} that for ${\rm m}={\rm f}\sqrt{2\lambda}$ and 
$\lambda\geqslant 0$: 
\begin{equation}
K(r)\sim e^{-Mr}\to 0, \quad H(r)\mp Mr \sim e^{-{\rm m}r}\to 0,\;\mbox{ i.e. }\;
S(r)\sim\frac{e^{-{\rm m}r}}{Mr}\to 0, 
\label{2_11} 
\end{equation}
as $r\to\infty$, where ${\rm m}\geqslant 0$ is associated with the mass of third Higgs boson 
after spontaneous symmetry breaking and quantisation over Higgs vacuum 
\cite{Grac,Rbk,Gd_Ol,Wpf,YS} (\ref{2_7}). 
 
The boundary conditions at $r\to 0$, preserving the finite monopole mass as its finite full 
energy \cite{Gd_Ol,Wpf} at rest (\ref{2_18_4}), can be written as:  
\begin{equation}
K(r)-1 \sim O(r), \quad H(r)\sim O(r), \;\mbox{ i.e. }\; S(r)\sim O(1).
\label{2_12}
\end{equation}
In order to preserve the regularity of the fields at the origin $r\to 0$ they should be 
imposed \cite{Blw,JZ,Grac,Rbk,YS} in a stronger form: 
\begin{equation}
K(r)-1 \sim O(r^2), \quad H(r)\sim O(r^2), \;\mbox{ i.e. }\; S(r)-1\sim O(r).
\label{2_12_1}
\end{equation}
The conditions (\ref{2_11}) and (\ref{2_12_1}) are fulfilled e.g. for known exact BPS- 
solution \cite{Rbk,Gd_Ol,Wpf,YS,Svrz,Bgm} in the limit  $\lambda\to 0$. However, for the 
solutions with $\lambda>0$ only the existense theorem \cite{Rbk,Grs,Zng}, but not the 
uniqueness theorem is known for both the conditions (\ref{2_11}) and (\ref{2_12}) or 
(\ref{2_12_1}) simultaneously. 
On the other hand, the conditions (\ref{2_11}) are enough to provide a finite abelian magnetic 
charge \cite{Ar,Blw,JZ,Grac,Rbk,Gd_Ol,Wpf,YS,Svrz}. 
So, the constraints (\ref{2_12}) or (\ref{2_12_1}) at $r=0$ do not have the same importance as 
the first one (\ref{2_11}) at $r\to\infty$ \cite{Ar,Rbk}, at least as long as we remain in 
the rest frame of the monopole field. The well known example of such kind is the 
non-abelian Wu-Yang monopole solution in $SU(2)$ Yang-Mills gauge theory \cite{YS,Svrz,Cho_13}. 
We will face the similar solutions below. 

\section{Other monopole solutions} 
Another possibility to realise the observed symmetry of the system 
(\ref{2_17_1})--(\ref{2_17_4}) is to impose $\beta=\eta Y$ with $\eta=\,$const. 
Then the Eq. (\ref{2_17_3}) gives $\eta^2=-1$. So we have two sets of complex classical 
solutions for every $\eta=\mp i$. Their complexification thus naturally arises from 
Euler equations without any changing \cite {Grac} of the Lagrangian (\ref{2_4}) and energy 
density (\ref{2_6}). So, the third equation of the system holds identically, whereas each of 
the first two Eqs. (\ref{2_17_1}), (\ref{2_17_2}) now coincides with the following:
\begin{eqnarray}
&&\!\!\!\!\!\!\!\!\!\!\!\!\!\!\!\!\!\! 
r^2\frac{d}{dr}\left(\frac{rY^\prime+\eta\alpha Y}{r}\right)=
Y\left(H^2-1\right)-\eta\alpha\left(rY^\prime+\eta\alpha Y\right).
\label{2_22_0} 
\end{eqnarray}
For the both cases, with real Higgs field i.e. real $H(r)$,  
it is convenient to parameterize complex solutions of (\ref{2_22_0}) by two real functions 
$N(r)$ and $\omega(r)$ (with the omitted indexes $\pm$ for $\eta=\mp i$) as 
\begin{equation}
Y_{\pm}(r)=N(r)e^{\pm i\omega(r)}, \quad\; \beta_{\pm}(r)=\mp iY_{\pm}(r), 
\;\;\mbox{ with }\;\;Y^2_{\pm}+\beta^2_{\pm}\equiv 0.
\label{2_22} 
\end{equation}
The Eq. (\ref{2_22_0}) for both solutions then reads as 
\begin{equation}
r^2N^{\prime\prime}-\left[\left(H^2-1\right)+\left(r\omega^\prime-\alpha\right)^2\right]N=
\mp i\left[2rN^\prime-N+Nr\frac{d}{dr}\right]\left(r\omega^\prime-\alpha\right).  
\label{2_23} 
\end{equation}
Obviously its right hand side now is real, while its left one is pure imaginary. So both 
of them turn to zero separately. To fulfill this it is enough for the second relation 
(\ref{2_20}) hold again as $r\omega^\prime=\alpha$, thus eliminating the dependence on indexes 
$\pm$ of functions $N$, $\omega$ as well as any gauge dependence of the obtained system of 
equations (the function $N$ was introduced only as real but not as positive):  
\begin{equation}
r^2N^{\prime\prime}-\left(H^2-1\right)N=0,\qquad 
r^2H^{\prime\prime}=\frac{\lambda}{e^2}\left[H^2-(Mr)^2\right]H.
\label{2_24} 
\end{equation}
Its explicit exact solution for both monopole $(+)$ and antimonopole $(-)$ cases, satisfies to 
boundary conditions (\ref{2_11}), (\ref{2_12}) for $H(r)$ and to asymptotic condition for 
$N(r)$ similar to that for $K(r)$ (\ref{2_11}), and has the following form: 
\begin{equation}
H(r)=\pm Mr\equiv \pm \zeta, \;\mbox{ whence }\; 
N(r)={\rm N}_0\left(\frac{2\zeta}{\pi}\right)^{1/2}K_{i\varrho}(\zeta), \;\mbox{ for } \;
\varrho=\frac{\sqrt{3}}{2}, 
\label{2_25} 
\end{equation}
where for $\zeta>0$ $K_{i\varrho}(\zeta)$ is the Macdonald function \cite{GR}, which remains 
the same and real for both imaginary indices $\pm i\varrho$. 
So for arbitrary dimensionless constant ${\rm N}_0>0$ the function $N(r)$ behaves as 
($\Gamma(z)$- is Euler gamma-function)
\begin{eqnarray}
&&\!\!\!\!\!\!\!\!\!\!\!\!\!\!\!\!\!\! 
N(r)\longmapsto\left\{\begin{array}{cc}{\rm N}_0 e^{-\zeta},\qquad \;\mbox{ with }\;
\zeta=Mr, & \mbox{ as }\; r\to\infty; \\
\displaystyle {\rm N}_0  \left(\frac{2\zeta}{\varrho\sinh\pi\varrho}\right)^{1/2}
\cos\left[Im\left\{\ln\Gamma(i\varrho)\right\}-\varrho \ln\left(\frac{\zeta}2\right)\right], & 
\mbox{ as }\; r\to 0.  
\end{array}
\right. 
\label{2_26} 
\end{eqnarray}
Similarly to the above real case (\ref{2_19}), the complex non abelian gauge field 
(\ref{2_13}) takes now the form with all three gauge-dependent {\bf ss}- components: 
\begin{eqnarray}
&&\!\!\!\!\!\!\!\!\!\!\!\!\!\!
A^j_{a\,\pm}(\xv)=\varepsilon^{jka}n^k \frac{\left(N e^{\pm i\omega}-1\right)}{er}-
n^jn^a\frac{\alpha(r)}{er}-\left(\delta^{ja}-n^jn^a\right)\frac{(\mp i)N e^{\pm i\omega}}{er}.
\label{2_27_0} 
\end{eqnarray} 
Corresponding complex ``chromo-magnetic'' field strength (\ref{2_16}) becomes 
\begin{eqnarray}
&&\!\!\!\!\!\!\!\!\!\!\!\!\!\!
B^j_{a\,\pm}(\xv)=\frac{n^j n^a}{er^2}-
\frac{\left(\delta^{ja}-n^j n^a\right)}{er^2}rN^\prime e^{\pm i\omega} -
\frac{\varepsilon^{jka}n^k}{er^2}(\mp i)rN^\prime e^{\pm i\omega}, 
\label{2_27} 
\end{eqnarray} 
where the first gauge-independent longitudinal part exactly reproduces the non-abelian 
Wu-Yang monopole solution \cite{YS}, while the two gauge-dependent transverse contributions 
exponentially disappear at $r\to\infty$ according to (\ref{2_26}). 
The complex structure of these contributions in (\ref{2_27}) in some sense is similar to real 
and imaginary axises of the complex plane. These ``axises'' are mutually exchanged by the 
shifting $\omega\mapsto\widetilde{\omega}\mp\pi/2$ in accordance with above symmetry of 
equations of motion. 

In view of (\ref{2_18_3}), (\ref{A_3}), (\ref{A_6}), the energy-momentum tensor densities are 
also reduced exactly to that of non-abelian Wu-Yang monopole, since in (\ref{2_18_3}), 
(\ref{A_2}) all the terms excluding the first one vanish all together on the solution 
(\ref{2_22}), (\ref{2_25}), (\ref{2_27}): 
\begin{equation}
\Theta^{00}_{\underline{m}}(r)=\frac{1}{2\,e^2 r^4}, \qquad \Theta^{jk}_{\underline{m}}(\xv)=
\left(\frac{\delta^{jk}}3-n^jn^k\right)\frac {1}{e^2r^4}+
\frac{\delta^{jk}}{6}\,\frac{1}{e^2r^4}. 
\label{2_28} 
\end{equation}

It is worthwhile to remind that the `t Hooft abelian projection of non-abelian monopoles 
solutions gives the same Dirac abelian monopole field for any one of them. It is a simple 
matter to show that, in accordance with topological nature \cite{Ar} of magnetic charge 
$g=\pm 4\pi/e$, the `t Hooft abelian fields strength tensor 
\cite{Blw,JZ,Grac,Rbk,Gd_Ol,Wpf,YS,Svrz} 
\begin{eqnarray}
&&\!\!\!\!\!\!\!\!\!\!\!\!\!\!
{\cal F}^{\mu\nu}=\frac{h_aG^{\mu\nu}_a}{\left(h_bh_b\right)^{1/2}}-
\frac{\varepsilon^{abc}h_a}{e\left(h_dh_d\right)^{3/2}}
\left(\widehat{D}^\mu\widehat{h}\right)_b\left(\widehat{D}^\nu\widehat{h}\right)_c, 
\;\;\mbox{ for }\;j,k,l=1,2,3,
\label{2_29}  \\
&&\!\!\!\!\!\!\!\!\!\!\!\!\!\!
{\cal F}^{j0}={\cal E}^j, \quad {\cal F}^{jk}=-\varepsilon^{jkl}{\cal B}^l,
\;\;\mbox{ gives }\;\;\vec{\cal B}(\xv)=\pm \frac{\nv}{e r^2}=\frac{g\xv}{4\pi r^3}, 
\label{2_30} 
\end{eqnarray} 
for {\sl arbitrary} functions $H \gtrless 0$ and $\alpha$, $\beta$, $Y$ in (\ref{2_8}), 
(\ref{2_13}), (\ref{2_16}), giving also the same energy-momentum tensor (\ref{2_28}) for the 
abelian Dirac monopole field (\ref{2_30}). Inspite of that the solutions (\ref{2_27}), as 
non-abelian Wu-Yang solution, for both abelian and non-abelian monopole lead to the 
identically divergent full energy (\ref{2_18_1}), (\ref{2_28}), such kind of fields are used 
in a number of modern field theoretical considerations \cite{Cho_13,Cho_14,H_Ng,Vr_Kr,Ed_Nak}.  

\section{Other dyon solutions} 
According to Julia and Zee \cite{JZ}, the temporal stationary {\bf ss}- component of gauge 
field, with the same 3D-space components (\ref{2_13}), appears in the chosen rest frame as 
\begin{eqnarray}
&&\!\!\!\!\!\!\!\!\!\!\!\!\!\!  
A^0_a(\xv)=n^a\frac{J(r)}r,\;\;\mbox{ leaving }\;\; 
\left(\widehat{D}^0\widehat{h}\right)_a=0,\;\;\mbox{ whence  }\;\; 
\left(\widehat{D}^j\widehat{E}^j\right)_a=0, 
\label{2_31} \\
&&\!\!\!\!\!\!\!\!\!\!\!\!\!\!
\mbox{for } \;\;  
\left(\widehat{D}^j\widehat{h}\right)_a=-n^jn^a\frac{d}{dr}\left(\frac{H}{er}\right)-
\left(\delta^{ja}-n^j n^a\right)\frac{HY}{er^2}-\varepsilon^{jka}n^k \frac{H\beta}{er^2},  
\label{2_32_H} \\
&&\!\!\!\!\!\!\!\!\!\!\!\!\!\!
\mbox{with } \;\;  
E^j_a=G^{j0}_a=-n^jn^a\frac{d}{dr}\left(\frac{J}{er}\right)-
\left(\delta^{ja}-n^j n^a\right)\frac{JY}{er^2}-\varepsilon^{jka}n^k \frac{J\beta}{er^2},  
\label{2_32} 
\end{eqnarray} 
as ``chromo-electric'' field. With the same ansatz (\ref{2_8}) for the Higgs fields this   
explicitly demonstrates Julia-Zee correspondense \cite{JZ} 
$A^0_a(\xv)\leftrightharpoons h_a(\xv)$, being just a consequence of (\ref{2_2}), (\ref{2_3}). 
The equations of motion (\ref{2_5}) give now the system of five equations:  
\begin{eqnarray}
&&\!\!\!\!\!\!\!\!\!\!\!\!\!\!\!\!\!\! 
r^2\frac{d}{dr}\left(\frac{rY^\prime+\alpha\beta}{r}\right)=
Y\left(Y^2+\beta^2+H^2-J^2-1\right)-\alpha\left(r\beta^\prime-\alpha Y\right),
\label{2_33_1} \\
&&\!\!\!\!\!\!\!\!\!\!\!\!\!\!\!\!\!\! 
r^2\frac{d}{dr}\left(\frac{r\beta^\prime-\alpha Y}{r}\right)=
\beta\left(Y^2+\beta^2+H^2-J^2-1\right)+\alpha \left(rY^\prime+\alpha\beta\right),
\label{2_33_2} \\
&&\!\!\!\!\!\!\!\!\!\!\!\!\!\!\!\!\!\! 
\beta\left(rY^\prime+\alpha\beta\right)=Y\left(r\beta^\prime-\alpha Y\right), 
\label{2_33_3} \\
&&\!\!\!\!\!\!\!\!\!\!\!\!\!\!\!\!\!\! 
r^2\frac{d^2H}{dr^2}=H\left\{\frac{\lambda}{e^2}\left[H^2-(Mr)^2\right]+
2\left(Y^2+\beta^2\right)\right\},  
\label{2_33_4}  \\
&&\!\!\!\!\!\!\!\!\!\!\!\!\!\!\!\!\!\! 
r^2\frac{d^2J}{dr^2}=2J(Y^2+\beta^2).
\label{2_33_5} 
\end{eqnarray}
The third Eq. (\ref{2_33_3}) remains unchanged as Eq. (\ref{2_17_3}) of previous system, 
as a direct consequence of chosen reference frame. Indeed, with the use of Eqs. (\ref{2_31}), 
(\ref{2_32}), (\ref{2_16}), for the momentum density of (\ref{2_6_0}) and for the 
corresponding full momentum one has:
\begin{eqnarray}
&&\!\!\!\!\!\!\!\!\!\!\!\!\!\!\!
\Theta^{0j}_d(\xv)\mapsto \left(\vec{E}_a\times\vec{B}_a\right)^j=
\varepsilon^{jkl}E^k_a B^l_a=\frac{2n^j}{e^2 r^4}J
\bigl[Y\left(r\beta^\prime-\alpha Y\right)-\beta\left(rY^\prime+\alpha\beta\right)\bigr], 
\label{2_34_0} \\
&&\!\!\!\!\!\!\!\!\!\!\!\!\!\!\!
\left(J=0\, \mbox{ for }\, \Theta^{0j}_m(\xv)\right), \; \quad
\mbox{whence }\;\,{\rm P}^j_{m,d}[\Theta]=\int d^3{\rm x}\,\Theta^{0j}_{m,d}(\xv)\longmapsto 0, 
\end{eqnarray}
in accordance with the chosen rest frame for both monopole and dyon cases. In the same terms
the energy functional corresponding to density (\ref{2_6}) now takes the form: 
\begin{eqnarray}
&&\!\!\!\!\!\!\!\!\!\!\!\!
{\rm E}_d[\Theta]=\frac{4\pi}{e^2}\int\limits^\infty_0\frac{dr}{r^2}\left\{
\frac 12\left[1-Y^2-\beta^2\right]^2+\left(rY^\prime+\alpha\beta\right)^2+
\left(r\beta^\prime-\alpha Y\right)^2+\right.
\nonumber \\ 
&&\!\!\!\!\!\!\!\!\!\!\!\!
\left.+\frac 12\left(rH^\prime-H\right)^2\!+\frac 12\left(rJ^\prime-J\right)^2\!+
\left(H^2+J^2\right)\left(Y^2+\beta^2\right)+
\frac{\lambda}{4e^2}\left[H^2-(Mr)^2\right]^2\!\right\}\!\!.  
\label{2_34_3} 
\end{eqnarray}
Corresponding stress tensor $\Theta^{jk}_d(\xv)$ is given by (\ref{A_1}), (\ref{A_2}) in 
Appendix. 
Similarly to monopole case the real substitution (\ref{2_19}) reduces the system 
(\ref{2_33_1})--(\ref{2_33_5}) and the energy (\ref{2_34_3}) to those obtained by Julia 
and Zee \cite{JZ,Rbk,Gd_Ol,Wpf,YS}, correspondingly as:   
\begin{eqnarray}
&&\!\!\!\!\!\!\!
r^2 K^{\prime\prime}=K\left[K^2+H^2-J^2-1\right], \qquad 
r^2 J^{\prime\prime}=2J K^2,
\label{2_35_1} \\
&&\!\!\!\!\!\!\!
r^2 H^{\prime\prime}=H\left\{\frac{\lambda}{e^2}\left[H^2-(Mr)^2\right]+2K^2\right\}, 
\label{2_35_2} 
\end{eqnarray}
and as (corresponding tensor $\Theta^{jk}_d(\xv)$ is given by (\ref{A_4}), 
(\ref{A_6})--(\ref{A_9}) in Appendix)
\begin{eqnarray}
&&\!\!\!\!\!\!\!
{\rm E}_d[\Theta]=\frac{4\pi}{e^2}\int\limits^\infty_0\frac{dr}{r^2}\left\{
\frac 12\left[1-K^2\right]^2+\left(rK^\prime\right)^2+\frac 12\left(rH^\prime-H\right)^2+
\right.
\nonumber \\
&&\!\!\!\!\!\!\!
\left. +\frac 12\left(rJ^\prime-J\right)^2+(H^2+J^2)K^2+
\frac{\lambda}{4e^2}\left[H^2-(Mr)^2\right]^2 \right\}. 
\label{2_35_3} 
\end{eqnarray} 
The reduction to these equations also takes place again for $\alpha=\beta=0$ and $Y\mapsto K$. 
In order to preserve the energy finite, the boundary conditions (\ref{2_11}), (\ref{2_12}) or 
(\ref{2_12_1}) should be supplemented now with the conditions \cite{JZ,Grac,Rbk,Gd_Ol} for the 
function $J(r)$: 
\begin{equation}
J(r)\to Cr+Q,\;\mbox{ as }\; r\to\infty,\;\mbox{ and }\; J(r)\sim O(r),\;\mbox{ or }\;
J(r)\sim O(r^2),\;\mbox{ as }\; r\to 0. 
\label{2_35_4} 
\end{equation}
The first of them, thanks to Eqs. (\ref{2_35_1}), changes also the asymptotic conditions 
(\ref{2_11}) for $K(r)$ as $r\to\infty$ to \cite{JZ}  
\begin{equation}
K(r)\sim e^{-r\sqrt{M^2-C^2}},\;\;\mbox{ and hence }\;\;J(r)-Cr-Q\sim e^{-2r\sqrt{M^2-C^2}}. 
\label{2_35_6} 
\end{equation}
Similarly to ``chromo-magnetic'' strength (\ref{2_21}), (\ref{2_21_1}), the ``chromo-electric' 
strength (\ref{2_32}) with real substitution (\ref{2_19}), becomes a gauge-dependent 
generalization (with $J\mapsto H$ for $(\widehat{D}^j\widehat{h})_a$ (\ref{2_32_H})) 
of Julia-Zee expression \cite{JZ}, corresponding here to $\omega=0$: 
\begin{equation}
E^j_a(\xv)=-n^jn^a\frac{d}{dr}\left(\frac{J}{er}\right)-
\Xi^{ja}(\nv,\omega) \frac{JK}{er^2}.  
\label{2_35_5} 
\end{equation}

Similarly to previuos section, the third Eq. (\ref{2_33_3}) here induces again the complex 
substitution $\beta=\eta Y$ with $\eta=\mp i$, which again reduces the first two Eqs. 
(\ref{2_33_1}), (\ref{2_33_2}) to one and the same equation for the complex function $Y(r)$: 
\begin{eqnarray}
&&\!\!\!\!\!\!\!\!\!\!\!\!\!\!\!\!\!\! 
r^2\frac{d}{dr}\left(\frac{rY^\prime+\eta\alpha Y}{r}\right)=
Y\left(H^2-J^2-1\right)-\eta\alpha\left(rY^\prime+\eta\alpha Y\right).  
\label{2_36_0} 
\end{eqnarray}
For the same complex substitution (\ref{2_22}) $Y_\pm=Ne^{-\eta\omega}$ this equation 
similarly transforms to 
\begin{equation}
r^2N^{\prime\prime}\!-\left[\left(H^2-J^2-1\right)+
\left(r\omega^\prime-\alpha\right)^2\right]\!N=
\mp i\left[2rN^\prime\!-N+Nr\frac{d}{dr}\right]\!\left(r\omega^\prime-\alpha\right).  
\label{2_37} 
\end{equation}
Here, for the same reasons as above, the second relation (\ref{2_20}) for real $J^2$ leads 
again to the gauge-independent system of three equations, instead of (\ref{2_35_1}), 
(\ref{2_35_2}):  
\begin{equation}
r^2 N^{\prime\prime}-\left(H^2-J^2-1\right)N=0,\quad \; 
r^2 H^{\prime\prime}=\frac{\lambda}{e^2}\left[H^2-(Mr)^2\right]H,\quad \;
r^2 J^{\prime\prime}=0. 
\label{2_38} 
\end{equation}
Its explicit exact solution similar to (\ref{2_25}) now reads (for real $C^2,Q^2,CQ$) as: 
\begin{eqnarray}
&&\!\!\!\!\!\!\!\!\!\!\!\!\!\!\!\!\!\! 
H(r)=\pm Mr,\quad \; J(r)=Cr+Q, \;\; \mbox{ whence, for }\;\;
\overline{\zeta}=r\sqrt{M^2-C^2}>0, 
\label{2_39} \\
&&\!\!\!\!\!\!\!\!\!\!\!\!\!\!\!\!\!\! 
\overline{\varrho}=\sqrt{\frac 34+Q^2},\quad \kappa=\frac{CQ}{\sqrt{M^2-C^2}},
\;\;\mbox{ one has }\;\; N_d(r)={\rm N}_0 W_{\kappa,i\overline{\varrho}}(2\overline{\zeta}), 
\label{2_40} 
\end{eqnarray}
where $W_{\kappa,i\overline{\varrho}}(z)$ is Whittaker function \cite{GR}.  
So for arbitrary dimensionless constant ${\rm N}_0>0$, the function $N_d(r)$ behaves 
similar to function $K(r)$ in (\ref{2_35_6}) as
\begin{eqnarray}
&&\!\!\!\!\!\!\!\!\!\!\!\!\!\!\!\!\!\! 
N_d(r)\longmapsto\left\{\begin{array}{cc}{\rm N}_0 \left(2\overline{\zeta}\right)^\kappa 
e^{-\overline{\zeta}}, & \mbox{ as }\; r\to\infty; \\  \displaystyle 
{\rm N}_0 \left|\frac{\Gamma(2i\overline{\varrho})}
{\Gamma(\frac 12-\kappa+ i\overline{\varrho})}\right|\left(2\overline{\zeta}\right)^{1/2}
\cos\left[\Phi_\kappa(\overline{\varrho})-\overline{\varrho}\ln(2\overline{\zeta})\right],& 
\mbox{ as }\; r\to 0,  
\end{array}
\right. 
\label{2_41} \\
&&\!\!\!\!\!\!\!\!\!\!\!\!\!\!\!\!\!\! 
\mbox{for }\;
\Phi_\kappa(\overline{\varrho})=Im\left\{\ln\left[\frac{\Gamma(2i\overline{\varrho})}
{\Gamma(\frac 12-\kappa+ i\overline{\varrho})}\right]\right\},\quad M^2-C^2>0,\;\mbox{ with }\; 
C^2,Q^2 \gtrless 0. 
\label{2_42}
\end{eqnarray}
For $C=0$, i.e. $\kappa=0$, $\overline{\zeta}\mapsto\zeta$, with \cite{GR}  
$W_{0,i\overline{\varrho}}(2\zeta)=\sqrt{2\zeta/\pi}\,K_{i\overline{\varrho}}(\zeta)$, 
this solution for the dyon function $N_d(r)$ reduces to the previous monopole one 
(\ref{2_25}), (\ref{2_26}) with $\varrho\mapsto \overline{\varrho}$. 

In terms of functions $N\mapsto N_d$ and $\omega$, the complex non-abelian gauge field 
(\ref{2_27_0}) and magnetic strength (\ref{2_27}) remain the same as for monopole. Their 
complex structure, as well as for the ``chromo-electric'' field strength (\ref{2_32}), 
now, instead of (\ref{2_21_1}), can be adsorbed into the complex transverse gauge-dependent 
tensor: 
\begin{equation}
\Upsilon^{ja}_{\pm}(\nv,\omega)=e^{\pm i\omega}\left[\left(\delta^{ja}-n^j n^a\right)
\mp i \varepsilon^{jka}n^k\right],\;\mbox{ with }\;
\frac{\Upsilon^{ja}_{+}+\Upsilon^{ja}_{-}}2=\Xi^{ja}(\nv,\omega).
\label{2_44} 
\end{equation}
Instead of real case (\ref{2_21}), (\ref{2_35_5}), the fields and strengths with (\ref{2_39}) 
can be written as 
\begin{eqnarray}
&&\!\!\!\!\!\!\!\!\!\!\!\!\!\!
A^j_{a\,\pm}(\xv)=-\,\frac{\varepsilon^{jka}n^k}{er}-n^jn^a\frac{\omega^\prime(r)}{e} 
\pm i\Upsilon^{ja}_{\pm} \frac{N_d(r)}{er}, \quad \;\;
\left(\widehat{D}^j_{\pm}\widehat{h}\right)_a=-\Upsilon^{ja}_{\pm} \frac{H N_d}{er^2},
\label{2_44_0} \\
&&\!\!\!\!\!\!\!\!\!\!\!\!\!\!
B^j_{a\,\pm}(\xv)=\frac{n^j n^a}{er^2}-\Upsilon^{ja}_{\pm} \frac{rN^\prime_d}{er^2}, 
\qquad  \;
E^j_{a\,\pm}=-n^jn^a\frac{d}{dr}\left(\frac{J}{er}\right)-
\Upsilon^{ja}_{\pm} \frac{JN_d}{er^2}.  
\label{2_44_1} 
\end{eqnarray}
According to (\ref{2_5}), (\ref{2_6_0}), (\ref{2_31}) they obey one and the same 
``self-dual-like'' relations:   
\begin{equation}
\left(\widehat{D}^j_{\pm}\widehat{E}^j_{\pm}\right)_a=0, \qquad 
\left(\widehat{D}^j_{\pm}\widehat{B}^j_{\pm}\right)_a=0, \qquad 
\left(\widehat{D}^j_{\pm}\widehat{D}^j_{\pm}\widehat{h}\right)_a=0. 
\label{2_46_0} 
\end{equation}

The obtained pairs of complex conjugate solutions (\ref{2_22}), (\ref{2_39}), (\ref{2_40}), 
(\ref{2_44})--(\ref{2_44_1}) belong to the complexified algebra $so(3,C)$ which has the 
same generators as its original real form \cite{Is_Rb,Lh_B} $so(3,R)\equiv so(3)$. 
They show that some of the symmetry restrictions that are imposed on the fields from the very 
beginning by the reasons of formal mathematic convenience in any case should be relaxed for 
the specific classical solutions, as a rule, due to their dynamical properties (e.g. the 
translation invariance already for the real original `t Hooft-Polyakov monopole solution 
(\ref{2_8}), (\ref{2_9})). 
Nevertheless, this relaxation may not destroy at all the symmetry structure of the model,  
but only change the way of its realisation \cite{YS}, e.g., as when instead of the one 
real solution (\ref{2_13}), (\ref{2_19}), (\ref{2_21}), (\ref{2_21_1}), \ref{2_35_5}) the 
above two complex conjugate ones appear. 
The field of specific classical solution itself is not the group parameter of gauge 
transformations, but it defines a specific equivalence class with respect to them. 
As shown in Appendix B, both equivalence classes for above real and complex original 
solutions ($\{0\}$ for $\omega=0$) are given by the one and the same automorphism of 
those different elements of the same algebra \cite{Is_Rb,Lh_B} $so(3,C)\approx so(3)$.
Both these different classes of spherically symmetric solutions are defined by the same 
spherically symmetric gauge transformation \cite{p_s,i_z} $U\left(\xv\right)\in SO(3)$, where 
$U\left(\xv\right)=\exp\left\{i\theta^a(\xv)T^a\right\}$ with $\theta^a(\xv)=n^a\omega(r)$, 
for arbirary real regular gauge function $\omega(r)$ (\ref{2_20}), as:  
\begin{eqnarray}
&&\!\!\!\!\!\!\!\!\!\!\!\!\!\!
\widehat{A}^{\{0\}}_\mu(\xv)\mapsto\widehat{A}^{\{\omega\}}_\mu(\xv)=
U(\xv)\left(\widehat{A}^{\{0\}}_\mu(\xv)+\frac{i}{e}\partial_\mu\right)U^{-1}(\xv),  \quad 
\widehat{A}^{\{\omega\}}_0(\xv)=\widehat{A}^{\{0\}}_0(\xv). 
\label{2_46_2} \\
&&\!\!\!\!\!\!\!\!\!\!\!\!\!\!
\widehat{G}^{\mu\nu}_{\{0\}}(\xv)\mapsto \widehat{G}^{\mu\nu}_{\{\omega\}}(\xv)=
U(\xv)\widehat{G}^{\mu\nu}_{\{0\}}(\xv)U^{-1}(\xv),\;\mbox{ i.e. }\;  
\widehat{\vec{B}}_{\{0\}}(\xv)\mapsto \widehat{\vec{B}}_{\{\omega\}}(\xv), 
\label{2_46_1} \\
&&\!\!\!\!\!\!\!\!\!\!\!\!\!\! 
\widehat{\vec{E}}_{\{0\}}(\xv)\mapsto \widehat{\vec{E}}_{\{\omega\}}(\xv), \quad 
\widehat{h}_{\{0\}}(\xv)\mapsto \widehat{h}_{\{\omega\}}(\xv)=
U(\xv)\widehat{h}_{\{0\}}(\xv)U^{-1}(\xv)=\widehat{h}_{\{0\}}(\xv),
\label{2_46_1_0} 
\end{eqnarray}
Thus, the relaxation of the symmetry restrictions concerns only the dynamical structure of the 
solutions, and not their gauge-dependent structure parameterising this gauge equivalence.  
So, the complex classical solutions to dynamical equations for $A^\mu_a$ or $G^{\mu\nu}_a$ 
do not mean here the complexification \cite{Is_Rb,Lh_B} of gauge group $SO(3)\equiv SO(3,R)$ 
to $SO(3,C)$ and do not destroy the gauge symmetry of the model. 


The abelian electric strength arises as a longitudinal field from (\ref{2_8}), 
(\ref{2_29})--(\ref{2_32}). 
For (\ref{2_39}), (\ref{2_44_1}) it is generated by point electric charge 
$q=Qg$ with magnetic $g=\pm 4\pi/e$
\begin{equation}
{\cal E}^j=\frac{h_a E^j_a}{\left(h_bh_b\right)^{1/2}}=
\mp \frac{n^j}e\frac{d}{dr}\left(\frac{J}r\right), \quad
\vec{\cal E}(\xv)=\pm \frac Qe\frac{\nv}{r^2}=Q\vec{\cal B}(\xv),\quad 
q=\!\int\! d^3{\rm x}\left(\vec{\nabla}\cdot {\vec{\cal E}}\right). 
\label{2_46} 
\end{equation}
Since the solution (\ref{2_39}) for $J(r)$ does not satisfy to boundary condition  
(\ref{2_35_4}) at the origin, the electric charge here can not be expressed as the known 
integrals  \cite{JZ,YS} over variable $r$ only, giving here formally only zero results:  
$q\mapsto 0$, if ($K^2\equiv Y^2+\beta^2$)
\begin{equation} 
q=\mp \frac{4\pi}e \int\limits^\infty_{0}rdr \frac{d^2 J}{dr^2}=
\mp \frac{8\pi}e \int\limits^\infty_{0}\frac{dr}r JK^2=-\,\frac{4\pi}{Me}
\int\limits^\infty_{0}dr\frac{d}{dr}\left\{rH \frac{d}{dr}\left(\frac{J}r\right)\right\}.
\label{2_47} 
\end{equation}
The reason is that for $\xv=r\nv$ the operator expression \cite{Kr_Li} 
$\vec{\nabla}_{\xv}\leftrightharpoons \nv\partial_r+(1/r)\vec{\partial}_{\nv}$ takes place 
only on the fields that are regular enough at $r=0$, or excluding this point.  

The obtained complex dyon solutions (\ref{2_22}), (\ref{2_39}), (\ref{2_40}), 
(\ref{2_44})--(\ref{2_44_1}), as well as the monopol ones (\ref{2_25})--(\ref{2_27}), are 
independent of $\lambda$ at all. 
Unlike  the (also generalized here) `t Hooft-Polyakov and Julia-Zee solutions (\ref{2_21}), 
(\ref{2_35_5}), they both additivelly contain the Wu-Yang-like  monopole solution as their 
exact longitudinal parts, that simmultaneously are their exact asymptotic items at 
$r\to\infty$ for both ``chromo-magnetic'' and ``chromo-electric'' fields:  
\begin{equation} 
B^j_{a\,\pm}(\xv)\underset{r\to\infty}{\longrightarrow}\frac{n^j n^a}{er^2},\qquad  
E^j_{a\,\pm}(\xv)\underset{r\to\infty}{\longrightarrow} Q\frac{n^j n^a}{er^2}
\approx Q B^j_{a\,\pm}(\xv).    
\label{2_48} 
\end{equation}                                
This approximate asymptotic relation  
transforms to the identity (\ref{2_46}) for abelian strengths corresponding to complex 
solutions (\ref{2_44_0}), (\ref{2_44_1}).
For $Q=\mp i$ it also means \cite{Grac} the approximate asymptotically self- or 
anti-self-dual non-abelian fields respectively. According to (\ref{2_41}), this approximate 
(anti) self-duality of the fields (\ref{2_44_1}) takes place for $r>R_0$ with the 
``self-duality radius'', defined by $R^{-1}_0=\sqrt{M^2-C^2}>0$. 

According to fixed boundary conditions at $r\to\infty$ (\ref{2_35_4}), (\ref{2_35_6}), the 
generalized Julia-Zee dyon solutions (\ref{2_21}), (\ref{2_35_5}) have the same 
asymptotic behavior (\ref{2_48}) with the same asymptotical self-duality relation, when 
$Q=\mp i$. 
The corrections to this relation, coming as (\ref{2_35_6}) from functions $K(r)$ and $J(r)$, 
are of the order of $e^{-r/R_0}$ for transversal parts of the field strengths and of the order 
of $e^{-2r/R_0}$ for their longitudinal parts. However, for the obtained here complex solution 
(\ref{2_44_1}) the corrections to longitudinal part are absent at all. 

The abelian strengths for generalized Julia-Zee dyon solutions (\ref{2_21}), (\ref{2_35_5})
are, of course, the same as 
given by Julia and Zee \cite{JZ}: there are the magnetic field (\ref{2_30}) of point-like 
Dirac monopole and electric field given by the first relation (\ref{2_46}) only. 
However, this electric field \cite{JZ,YS} here is not the field of point-like electric charge 
from the second Eq. (\ref{2_46}). This equation becomes now only the approximate asymptotic 
relation between the abelian electric and magnetic strengths, and the full electric charge is 
given here by Eqs. (\ref{2_47}).  

On the solutions (\ref{2_39})--(\ref{2_44_1}) the energy-momentum tensor (\ref{2_34_3}), 
(\ref{A_2}), (\ref{A_3}), (\ref{A_6}) coinsides again with the abelian one for the fields 
(\ref{2_46}) and can be written as 
\begin{equation}
\Theta^{00}_{\underline{d}}(r)=\frac{(1+Q^2)}{2\,e^2r^4}, \qquad 
\Theta^{jk}_{\underline{d}}(\xv)=\left(\frac{\delta^{jk}}3-n^jn^k\right)\frac{(1+Q^2)}{e^2r^4}+
\frac{\delta^{jk}}6 \frac{(1+Q^2)}{e^2r^4}. 
\label{2_45} 
\end{equation}
Thus, the obtained complex solutions have real energy-momentum density $\Theta^{\mu\nu}(\xv)$, 
but their full energies divergent at the origin similar to Wu-Yang solution. At the same time, 
the expressions (\ref{2_45}) represent the asymptotic form of energy-momentum tensor of 
Julia-Zee solution at $r\to\infty$. 
Indeed, according to (\ref{2_11}), (\ref{2_35_6}), at least for $\lambda>0$, the corrections 
to this asymptotic form of $\Theta^{00}_{d}(r)$ in (\ref{2_35_3}) and $\Theta^{jk}_{d}(\xv)$ 
(\ref{A_4}) are, evidently, of exponential type only, of the orders of $e^{-2r/R_0}$ 
and $e^{-2{\rm m}r}$.  

With the help of corresponding embedding \cite{Is_Rb,Lh_B} of algebra $so(3)$, 
the similar solutions may be obtained for the more wide gauge groups 
\cite{YS,Svrz,MPr}. 

\section{Conclusions and outlook}
The parity symmetry arguments \cite{Rbk}, as (4.13) of Goddard and Olive \cite{Gd_Ol}: 
$$
A^j_a(-\xv)\mapsto -A^j_a(\xv), \quad A^0_a(-\xv)\mapsto A^0_a(\xv),
\quad h_a(-\xv)\mapsto h_a(\xv),
$$
which reduce the potential (\ref{2_13}) to the `t Hooft ansatz (\ref{2_8}), do not seem to be 
well justified for the configuration and isotopic spaces simultaneously (especially if one 
compare their ansatz (4.15) to (4.13)), because the meaning of ``parity symmetry'' in 
isotopic (color) space is not well defined \cite{Svrz,Corrig}.  
In fact, here we deal simply with the additional Coulomb and transverse gauge conditions 
removing the functions $\alpha(r)$ and $\beta(r)$ according to Eqs. (\ref{2_14}). 
We have shown here that the removing of that constraints can be described in analytical way.  
It leads to explicit generalization of the known `t Hooft-Polyakov and Julia-Zee solutions for 
arbitrary gauge and gives the new complex $\lambda$- independent solutions (\ref{2_39}), 
(\ref{2_40}), (\ref{2_44_0}), (\ref{2_44_1}) for non-abelian monopole and dyon fields, 
that have simple real energy-momentum tensor (\ref{2_45}) and are tightly connected with the 
known non-abelian Wu-Yang solution \cite{Wpf,YS}. 

Nevertheless, one can also obtain zero energy-momentum tensor (\ref{2_45}) i.e. solution with 
zero mass and stress, choosing the parametres of complex non-abelian solution (\ref{2_39}), 
(\ref{2_40}), (\ref{2_45}) for real $J^2$ as $Q=\mp i$, with $Re\,C=0$. This, at first sight
exotic, formal choice corresponds to exact (anti) self-duality (\ref{2_48}) of longitudinal 
parts of the strengths and delivers the absolute minimum to the energy (\ref{2_34_3}), 
(\ref{2_45}) of the solution as ${\rm E}_{\underline{d}}[\Theta]=0$. 
For this choice, $i\overline{\varrho}=\pm 1/2$, and this solution describes massless complex 
self-dual dyon saturating the Bogomol'nyi bound \cite{YS,Bgm} 
${\rm E}_{d}[\Theta]\geqslant{\rm f}\sqrt{g^2+q^2}$ by real magnetic charge $g$, equipped with 
imaginary ``electric'' charge $q=Qg$ with $Q^2=-1$. For abelian strengths this corresponds 
formally to pure imaginary ``electric field'' (\ref{2_46}) of such complex dyon. 
The remaining arbitrary imaginary parameter $C$ of this self-dual dyon solution governs 
the value of its self-duality radius $R_0$. E.g. for $C=0$, the asymptotics (\ref{2_41}) at 
$r\to\infty$ becomes a simple exact solution for 
$N_{d}(r)= {\rm N}_0 W_{0,1/2}(2\zeta)={\rm N}_0 e^{-\zeta}$ with $R_0=1/M$, 
similar to (\ref{2_26}).  
For $C\to\pm i\infty$ independently of $Q$, with $\kappa\mapsto \pm iQ$, there follows  
from the asymptotics (\ref{2_41}), that 
$N_d(r)={\rm N}_0 W_{\kappa,i\overline{\varrho}}(2\overline{\zeta})\mapsto 0$ as 
$\overline{\zeta}\to +\infty$, giving $R_0=0$. So the non-abelian Wu-Yang monopole 
solution is reproduced exactly by magnetic strength (\ref{2_44_1}) for this choice of 
parameter $C$, and not only asymptotically as $r\to\infty$ in (\ref{2_48}). 
In other words, for $Q=\mp i$ with that choice of $C$, the (anti) self-duality becomes 
an exact feature for both non-abelian and abelian strengths (\ref{2_44_1}), (\ref{2_46}) 
simultaneously. 

On the other hand, the case $Y=\beta=0$ leads to the same non-zero exact dyon solution of 
Georgi-Glashow model, given by Eqs. (\ref{2_39}), (\ref{2_40}), (\ref{2_44_0}), (\ref{2_44_1}) 
for ${\rm N}_0=0$. 
So, it is also given by the relations (\ref{2_48}), becoming the exact equalities. 
For real $Q$, it may also be considered formaly as a real particular solution of Julia-Zee 
equations (\ref{2_35_1}), (\ref{2_35_2}) with $K(r)\equiv 0$. 
However, it becomes again (anti) self-dual and massless for $Q=\mp i$. 
So, for classical solutions to non-abelian equations of motions for $A^\mu_a$ 
there are at least two independent sources of complexity: from $A^0_a(\xv)$ and/or 
$A^j_a(\xv)$. 

The discussed massless complex (anti) self-dual dyon solutions can have a particular interest 
e.g. as a possible carriers of dark energy \cite{Gr_Rb,Lk_Rb,Chrn}. Indeed, due to their 
masslessness it is impossible to connect any Lorentz reference frame to them
\cite{Gr_Rb,Chrn}. 
Thanks to the pure imaginary value of corresponding interaction forces, their imaginary 
``electric'' charges by definition should be fully invisible for usual electric and 
magnetic charges and currents, and thus for usual Maxwell electromagnetic fields. 
Furthermore, since the usual charged pairs creation is a pure electric effect \cite{i_z} with 
probability proportional to $\vec{\cal E}^2$, it is then impossible for pure imaginary 
``electric'' charges and fields (\ref{2_46}) with $Q=\mp i$. 
That means also the absence for them of the usual loop corrections \cite{i_z,Grib}, as well as 
of the known Gribov confinement \cite{Grib} of massless electric charges. Moreover, for all 
known classical, quantum mechanical and quantum field cases \cite{Schw,Schw_m,Milt,YS}, the 
relative abelian dynamic of two such, even complex, dyons in their center of mass frame is 
described by two combinations of their charges: 
$$
{\cal C}=g_1g_2+q_1q_2=g_1g_2(1+Q_1Q_2),\;\mbox{ and }\; 
{\cal M}=q_1g_2-q_2g_1=g_1g_2(Q_1-Q_2). 
$$
Thus, when $Q_1=Q_2$ for $Q^2 _{1,2}=-1$, both the ``Coulomb'' hermitian ${\cal C}$ and the 
``magnetic'' non hermitian ${\cal M}$ contributions exactly disappear. That means that for two 
such complex dyons, for the same sign of their real magnetic charges $g_{1,2}$ in (\ref{2_30}), 
their mutual magnetic repulsion may be either twiced in ${\cal C}/r_{12}$ by respective 
Coulomb ``electric'' interaction of the imaginary `electric'' charges with opposite signs or 
exactly compensated by this interaction for the dyons with the {\sl same sign} of ``electric'' 
charges. 
For the same reasons, the mutual attraction of two real magnetic charges $g_{1,2}$ with 
opposit signs in (\ref{2_30}) also may be either twiced or exactly compensated by Coulomb 
interaction of imaginary ``electric'' charges of those complex self-dual dyons. For the both  
last cases of compensation with ${\cal C}=0$, the value ${\cal M}=0$ automatically, and the 
motion of self-dual dyons becomes completely free. 

Therefore, the filling of the vacuum with those complex self-dual dyons with the one and the 
same sign of their real magnetic charges and with the one and the same sign of their imaginary 
``electric'' charges, with constant particle $\overline{n}_\upsilon$ and energy 
$\overline{\rho}_\upsilon$ densities, naturally gives the necessary linear with distance 
macroscopic effect of antigravity for usual matter \cite{Gr_Rb,Lk_Rb,Chrn}. 
So, being completely unobservable, such complex dyons can serve as effective carriers of dark 
energy as a kinetic energy of their free motion with speed of light, manifestating in 
the constant energy density $\overline{\rho}_\upsilon$ of that ``dyonic vacuum medium''. 
Considering this density as ``equilibrium'' energy density of the spinless \cite{YS} complex 
dyon ``gas'', with the conventional value \cite{Gr_Rb} 
$\overline{\rho}_\upsilon\sim 10^{-46}(Gev)^4$, one finds 
$
\overline{n}_\upsilon\approx
1,2\,\pi^{-2}\left(30\,\pi^{-2}\overline{\rho}_\upsilon \right)^{3/4}\sim 10^6\, cm^{-3}, 
$
as estimation for corresponding ``equilibrium'' particle density.  
Of course, this is very rough estimate, because the pressure of that ``vacuum medium''  
should be \cite{Gr_Rb,Lk_Rb,Chrn} $p_\upsilon=-\overline{\rho}_\upsilon$, and not as for the 
free equilibrium gas with $p=\overline{\rho}/3$. 

Besides, any simply connected compact macroscopic domain of such a ``dyonic vacuum medium'' 
with only the opposite sign of imaginary ``electric'' charges will ``float up'' as a whole in 
the first one also with a force, obviously linear with distance. 
Perhaps this can be used to explain different inflation rates at different scales 
\cite{Gr_Rb, Lk_Rb}. 
Here it is also natural to suppose, that contributions of non hermitian ``magnetic'' forces 
proportional to ${\cal M}$ vanish on macroscopic averaging over their directions.
(For real dyons they disappear as pure gauge since the value of ${\cal M}$ is quantised 
\cite{Schw,Schw_m,Milt,YS}.)

On the other hand, the gauge-independent potential and strength decompositions, with exactly 
isolated non-abelian Wu-Yang-monopole-like contributions obtained here for the solutions 
(\ref{2_44_0}), (\ref{2_44_1}), are similar to gauge-independent decompositions of these 
quantities obtained in QCD \cite{Cho_13,Cho_14}. They are used there for explanation of 
monopoles condensation, leading to color confinement due to dual Meissner effect \cite{Cho_13}. 
Since the non abelian monopole and antimonopole are physically undistinguishable in QCD 
\cite{Cho_13}, they should have the same abelian magnetic charges $g$. 
Thus, if the last QCD decompositions \cite{Cho_13,Cho_14} can be reduced to the first 
ones $\lambda$- independent (\ref{2_39}), (\ref{2_44_0}), (\ref{2_44_1}) together with the 
self-duality relations like (\ref{2_46}), (\ref{2_48}), also with pure imaginary values of 
abelian ``electric'' charge for $Q=\mp i$, then the both types of ``vacuum media'' with the 
massless complex self-dual dyons, those for dark energy and for color confinement, 
will coincide.  
\appendix 
\section{}
For stationary spherically symmetric case (\ref{2_8}), (\ref{2_31}) the stress tensor is 
defined by (\ref{2_6_0}), (\ref{2_7}) as follows:  
\begin{eqnarray}
&&\!\!\!\!\!\!\!\!\!\!\!\!\!\!\!\!\!\! 
\Theta^{jk}(\xv)=
\delta^{jk}\left\{\frac 12\left[\left(\vec{E}_a\right)^2+\left(\vec{B}_a\right)^2\right]-
\frac 12\left[\left(\widehat{D}^l\widehat{h}\right)_a\right]^2-
\frac{\lambda}{4}\left(h_ah_a-{\rm f}^2\right)^2\right\}-
\nonumber \\ 
&&\!\!\!\!\!\!\!\!\!\!\!\!\!\!\!\!\!\! 
-E^j_a E^k_a-B^j_a B^k_a+ 
\left(\widehat{D}^j\widehat{h}\right)_a \left(\widehat{D}^k\widehat{h}\right)_a.  
\label{A_1} 
\end{eqnarray}
Substitution of Higgs fields (\ref{2_8}) and strengths (\ref{2_16}), (\ref{2_32_H}), 
(\ref{2_32}) transforms it to:   
\begin{eqnarray}
&&\!\!\!\!\!\!\!\!\!\!\!\!\!\!\!\!\!\! 
\Theta^{jk}_d(\xv)=\left(\frac{\delta^{jk}}2-n^jn^k\right)
\left\{\left[\frac{1-Y^2-\beta^2}{er^2}\right]^2+
\left[\left(\frac{J}{er}\right)^\prime\right]^2-
\left[\left(\frac{H}{er}\right)^\prime\right]^2 \right\}+
\nonumber \\ 
&&\!\!\!\!\!\!\!\!\!\!\!\!\!\!\!\!\!\! 
+\frac{n^jn^k}{e^2r^4}\left\{\left(rY^\prime+\alpha\beta\right)^2+
\left(r\beta^\prime-\alpha Y\right)^2+\left(J^2-H^2\right)\left(Y^2+\beta^2\right)\right\}-
\nonumber \\ 
&&\!\!\!\!\!\!\!\!\!\!\!\!\!\!\!\!\!\! 
- \delta^{jk}\frac{\lambda}{4\,e^4r^4}\left[H^2-(Mr)^2\right]^2. 
\label{A_2} 
\end{eqnarray}
For the complex solution the substitution (\ref{2_22}), (\ref{2_39}), (\ref{2_40}) 
immediately reduces this to expression (\ref{2_45}) in the following simple form, explicitly 
showing its conservation: 
\begin{equation}
\Theta^{jk}_{\underline{d}}(\xv)=\left(\frac{\delta^{jk}}2-n^jn^k\right)\frac{(1+Q^2)}{e^2r^4}, 
\qquad \partial_j \Theta^{jk}_{\underline{d}}(\xv)=0, \;\;\mbox{ for }\; r\neq 0.  
\label{A_3}
\end{equation}
For $Q^2=0$ it is reduced to the corresponding monopole stress tensor (\ref{2_28}). 

The real substitution (\ref{2_19}) simplifies tensor (\ref{A_2}) to the following one 
\begin{eqnarray}
&&\!\!\!\!\!\!\!\!\!\!\!\!\!\!\!\!\!\! 
\Theta^{jk}_d(\xv)=\left(\frac{\delta^{jk}}2-n^jn^k\right)
\left\{\left[\frac{1-K^2}{er^2}\right]^2+
\left[\left(\frac{J}{er}\right)^\prime\right]^2-
\left[\left(\frac{H}{er}\right)^\prime\right]^2 \right\}+
\nonumber \\ 
&&\!\!\!\!\!\!\!\!\!\!\!\!\!\!\!\!\!\! 
+\frac{n^jn^k}{e^2r^4}\left\{\left(rK^\prime\right)^2+
\left(J^2-H^2\right)K^2\right\}-\delta^{jk}\frac{\lambda}{4\,e^4r^4}\left[H^2-(Mr)^2\right]^2,  
\label{A_4} 
\end{eqnarray}
corresponding to Julia-Zee solution \cite{JZ}. For $J=0$ it is reduced to monopole stress 
tensor $\Theta^{jk}_m(\xv)$ corresponding to `t Hooft-Polyakov solution \cite{Pol,tH}.  

The given form of Eqs. (\ref{2_28}), (\ref{2_45}) reflects the fact that any symmetric tensor 
$\Theta^{jk}$, e.g. $\Theta^{jk}(\xv)=\delta^{jk}u(r)+n^jn^k\sigma(r)$, may be identically 
decomposed into the mutually orthogonal traceless and diagonal parts as    
\begin{eqnarray}
&&\!\!\!\!\!\!\!\!\!\!\!\!\!\!\!\!\!\! 
\Theta^{jk}\equiv \left(\Theta^{jk}-\frac{\delta^{jk}}3 Tr\{\Theta\}\right)+
\frac{\delta^{jk}}3 Tr\{\Theta\},\;\; \mbox{ what gives } 
\label{A_5} \\
&&\!\!\!\!\!\!\!\!\!\!\!\!\!\!\!\!\!\! 
\Theta^{jk}(\xv)=\left(n^jn^k-\frac{\delta^{jk}}3\right)\sigma(r)+\delta^{jk}\Pi(r), 
\;\; \mbox{ for }\;\;\Pi(r)=u(r)+\frac 13 \sigma(r).
\label{A_6}
\end{eqnarray} 
According to M. Polyakov \cite{MVP,JMP}, $\Pi(r)$ represents the distribution of pressure, 
whereas $\sigma(r)$ gives the distribution of shear forces. 

For the complex dyon solution from (\ref{A_3}), (\ref{A_6}), in view of (\ref{2_45}), it 
follows that 
\begin{equation} 
e^2 r^4 \sigma_{\underline{d}}(r)=-\left(1+Q^2\right), \qquad \;
e^2 r^4 \Pi_{\underline{d}}(r)=\frac 16 \left(1+Q^2\right). 
\label{A_10}
\end{equation}
From Eq. (\ref{A_4}) for Julia-Zee dyon it follows, that:
\begin{eqnarray}
&&\!\!\!\!\!\!\!\!\!\!\!\!\!\!\! 
e^2 r^4 u_d(r)=\frac 12 \left[\left(1-K^2\right)^2\!+\left(rJ^\prime\!-J\right)^2\!-
\left(rH^\prime\!-H\right)^2\right]-\frac{\lambda}{4e^2}\left[H^2-(Mr)^2\right]^2,
\label{A_8}  \\
&&\!\!\!\!\!\!\!\!\!\!\!\!\!\!\!
e^2 r^4 \sigma_d(r)=\left(rK^\prime\right)^2\!+\left(J^2\!-H^2\right)\!K^2\!-
\left[\left(1-K^2\right)^2\!+\left(rJ^\prime\!-J\right)^2\!-
\left(rH^\prime\!-H\right)^2\right]\!, 
\label{A_7} \\
&&\!\!\!\!\!\!\!\!\!\!\!\!\!\!\! 
e^2 r^4 \Pi_d(r)=\frac 16 \left[\left(1-K^2\right)^2\!+\left(rJ^\prime\!-J\right)^2\!-
\left(rH^\prime\!-H\right)^2\right]-\frac{\lambda}{4e^2}\left[H^2-(Mr)^2\right]^2+
\nonumber \\
&&\!\!\!\!\!\!\!\!\!\!\!\!\!\!\! 
+\frac 13 \left[\left(rK^\prime\right)^2+\left(J^2-H^2\right)K^2 \right]. 
\label{A_9}
\end{eqnarray}
The analysis of stability conditions corresponding to these values for both 
Julia-Zee and `t Hooft-Polyakov solutions was done recently in the work of Iu. Panteleeva 
\cite{JP}. 
\section{}
The infinitesimal versions of the first of  Eqs. (\ref{2_46_2})   
\begin{eqnarray}
&&\!\!\!\!\!\!\!\!\!\!\!\!\!\!\! 
A^{a\{0\}}_\mu(\xv)\mapsto A^{a\{\omega\}}_\mu(\xv)=A^{a\{0\}}_\mu(\xv)+
\varepsilon^{abc}A^{b\{0\}}_\mu(\xv)\theta^c(\xv)+\frac{1}{e}\partial_\mu\theta^a(\xv), 
\label{B_3}
\end{eqnarray}
immediately coincide with that for (\ref{2_13}), (\ref{2_20}), (\ref{2_19}) and (\ref{2_44_0})  
respectively, with the one and the same infinitesimal $\theta^a(\xv)=n^a\omega(r)$. 
According to (\ref{2_8}), (\ref{2_31}) for the vector-matrix 
$\vec{\cal T}=\ev_j\delta^{ja}T^a$ defined in configuration space with 
Cartesian basis $\ev_j$, by writing 
$\theta^a(\xv)T^a=\omega(r)\left(\nv\cdot\vec{\cal T}\right)$ and 
$U(\xv)=\exp\left\{i\,\omega(r)\left(\nv\cdot\vec{\cal T}\right)\right\}$, one can get the 
other relations (\ref{2_46_2}), (\ref{2_46_1}), (\ref{2_46_1_0}) from the known 
rotation formulae \cite{Is_Rb,Lh_B}, that for arbitrary fixed vectors $\nv$, $\sv$, with  
$[\nv,T^a]=[\sv,T^a]=0$ and algebra (\ref{2_1}) look as: 
\begin{eqnarray}
&&\!\!\!\!\!\!\!\!\!\!\!\!\!\!\! 
e^{i\omega\left(\nv\cdot\vec{\cal T}\right)}\left(\nv\cdot\vec{\cal T}\right) 
e^{-i\omega\left(\nv\cdot\vec{\cal T}\right)}=\left(\nv\cdot\vec{\cal T}\right), \;\; 
\mbox{ but } \;\; 
e^{i\omega\left(\nv\cdot\vec{\cal T}\right)}\left(\sv\cdot\vec{\cal T}\right) 
e^{-i\omega\left(\nv\cdot\vec{\cal T}\right)}=
\label{B_1} \\
&&\!\!\!\!\!\!\!\!\!\!\!\!\!\!\! 
=\left(\sv\cdot\vec{\cal T}\right)\cos\omega+
\left(\sv\cdot\nv\right)\left(\nv\cdot\vec{\cal T}\right)(1-\cos\omega)+
\left((\sv\times\nv)\cdot\vec{\cal T}\right)\sin\omega,
\nonumber \\
&&\!\!\!\!\!\!\!\!\!\!\!\!\!\!\! 
\mbox{whence: }\;\;
\ev_j\Upsilon^{ja}_{\pm}(\nv,\omega)T^a\equiv \widehat{\vec{\Upsilon}}_{\pm}^{\{\omega\}}=
e^{i\omega\left(\nv\cdot\vec{\cal T}\right)}\widehat{\vec{\Upsilon}}_{\pm}^{\{0\}} 
e^{-i\omega\left(\nv\cdot\vec{\cal T}\right)},
\label{B_2}
\end{eqnarray}
and obviously the same for $\Xi^{ja}(\nv,\omega)$ (\ref{2_21_1}), (\ref{2_44}).

\section*{Acknowledgments}
Authors thank M. Polyakov, A. Rastegin, Yu. Markov, E. Aman, S. Lovtsov, A. Kalo\-shin for 
useful discussions, Ya. Shnir and Reviewer for important comments. 

\end{document}